\documentclass[12pt]{article}
\usepackage{graphicx}
\usepackage{latexsym}
\usepackage{amssymb}
\usepackage{amsmath}
\usepackage[textwidth=6.5in,textheight=9in]{geometry}

\usepackage{helvet}

\begin{document}

\fontfamily{phv}\selectfont{

\noindent\textbf{Magnetic field strengths of hot Jupiters from signals of star-planet interactions}
\bigskip

\noindent \textbf{Authors:} P. Wilson Cauley$^{1\star}$, Evgenya L. Shkolnik$^2$, Joe Llama$^3$, 
and Antonino F. Lanza$^4$

\bigskip

\noindent\textbf{Affiliations:} $^1$University of Colorado Boulder, Laboratory of Atmospheric and
Space Sciences, Boulder, CO, USA, $^2$Arizona State University, School of Earth
and Space Exploration, Tempe, AZ, USA, $^3$Lowell Observatory, Flagstaff, AZ,
USA, $^4$INAF-Osservatorio Astrofisico di Catania, Catania, Italy,$^\star$email: paca7401@lasp.colorado.edu

\bigskip

\textbf{Evidence of star-planet interactions in the form of planet-modulated
chromospheric emission has been noted for a number of hot Jupiters. Magnetic
star-planet interactions involve the release of energy stored in the stellar
and planetary magnetic fields. These signals thus offer indirect detections of
exoplanetary magnetic fields. Here we report the derivation of the magnetic
field strengths of four hot Jupiter systems using the power observed in Ca II K
emission modulated by magnetic star-planet interactions. By approximating the
fractional energy released in the Ca II K line we find that the surface
magnetic field values for the hot Jupiters in our sample range from 20 G to 120
G, $\approx 10-100$ times larger than the values predicted by dynamo scaling
laws for planets with rotation periods of $\approx 2 - 4$ days.  On the other
hand, these value are in agreement with scaling laws relating the magnetic
field strength to the internal heat flux in giant planets. Large planetary
magnetic field strengths may produce observable electron-cyclotron maser radio
emission by preventing the maser from being quenched by the planet's
ionosphere. Intensive radio monitoring of hot Jupiter systems will help confirm
these field values and inform on the generation mechanism of magnetic fields
in this important class of exoplanets.}

The close orbits of hot Jupiters ($a \lesssim 10 R_*$ where $a$ is the
planet's semimajor axis and $R_*$ is the stellar radius) make it possible for
these objects to experience strong interactions with their host stars via
tides, collisions with stellar wind particles, accretion of evaporating
planetary gas onto the stellar surface, and magnetic reconnection between the
stellar and planetary magnetic field lines. Star-planet interactions (SPI)
can potentially reveal details about stellar wind properties, for which there
are very little data$^1$, and, most intriguingly, planetary magnetic
fields$^{2,3,4,5,6}$. 

Magnetic fields play a critical role in shielding planetary atmospheres from
incoming stellar wind particles and limiting mass loss$^{7,8,9}$, a key
condition for habitability, and provide information about the composition of
planetary interiors$^{10}$. Understanding the magnetic fields of hot Jupiters
will aid in the future detection and characterization of the magnetic fields of
smaller planets closer to the habitable zone$^{11}$. Most critically, measuring
exoplanet magnetic field strengths may break the composition degeneracy for
planets with known mass and radius, providing information about the structure
and dynamics of the convecting and electrically conducting interior$^{10}$.

Magnetic SPI signals can manifest as flux changes in the cores of
chromospherically active lines that vary on the same timescale as the planet's
orbital period, unlike tidal interactions which appear on timescales of half
the orbital period$^2$.  Evidence of such modulations has been observed in a
number of hot Jupiter systems in the chromospheric emission of Ca II.  The
first such signal was presented for HD 179949$^{12}$.  Followup observations
demonstrated similar behavior, strengthening the magnetic SPI
interpretation$^{13,14}$.  Similar signals have been observed in HD 189733,
$\tau$ Boo, and $\upsilon$ And$^{15,16}$.  A flare signature in excess of
phased rotational variations has also been noted in HD 73256$^{13}$.  SPI-like
modulations have also been observed in optical photometric variations$^{17,18}$
and in X-rays$^{19,20,21}$. The observed variable nature of some magnetic SPI
signatures can be attributed to the planet passing through regions of differing
stellar magnetic field strength and topology$^{22,23,24}$.

Chromospheric line flux changes due to magnetic SPI may be used to estimate
planetary magnetic field strengths. Scaling laws relating the power in magnetic
SPI to planetary and stellar properties have been given by various
authors$^{2,5,25,26}$.  These power estimates are consistent for a variety of
different coronal field cases, although the exact constants change depending on
the assumptions made. It was demonstrated that, in the case of a dipolar
stellar field, the alignment of the stellar and planetary magnetospheres can
have a significant impact on the energy dissipated in the magnetic
interaction$^5$. An additional source of energy in SPI signals may also be
provided by particle outflows from the planet onto the stellar surface$^{27}$.
It has also been shown that the decrease of the relative helicity of the
stellar field, triggered by the close-in planet, allows the evolution of the
stellar magnetic field configuration towards a lower energy state thus making
dissipation of its magnetic energy by internal reconnection possible on a
global scale$^{25,26}$. This process can release a power up to 2-3 orders of
magnitude larger than a pure reconnection at the boundary between the stellar
and planetary magnetospheres.

With the exception of a power estimate for the HD 179949 SPI signal and the
tentative SPI deviations for HD 73256 and $\kappa^1$ Cet$^{13}$, no other SPI
measurements have been flux-calibrated. This is due to the fact that
ground-based high-resolution spectra are difficult to flux calibrate accurately
and, in general, absolute fluxes are not necessary to confirm a SPI detection.
However, in order to constrain specific SPI models or calculate absolute
strengths of exoplanetary magnetic fields, measurements of the power emitted in
SPI signals are needed$^{2,28}$. Exoplanet magnetic field strengths derived
from SPI signals offer an important comparison for future detections, e.g.,
from radio observations of electron-cyclotron maser emission with the
Low-Frequency Array (LOFAR)$^{29}$.

In this paper we present magnetic field strengths for a sample of hot Jupiters
with published SPI signals (see Table 1). Using a new flux-calibration method
for high-resolution spectra, we convert the observed SPI signals into emitted
powers and estimate the total power in the SPI mechanism by applying specific
models of magnetic SPI.  We find that the field strengths are larger than
predicted by rotation scaling relations for planets with rotation periods of
$\approx 2 - 4$ days$^{3,30,31}$, as is expected for tidally-locked hot Jupiters,
and are comparable to the values suggested by internal heat flux scalings
relationships which also account for the extra energy deposition in hot
Jupiters. Our results suggest that the heat flux description is the most
promising of the models considered for predicting the magnetic fields of hot
giant planets.

\bigskip

\noindent\textbf{Ionized calcium SPI signals.} We used existing data sets for
which the changes in the Ca II K (3933.66 \AA) cores have been attributed to
SPI due to modulation that is phased with the planet's orbital period
and exclude those for which the variability is attributable to stellar
rotation$^{15}$ (see Methods for details on the observations). Previously
published SPI signals are not, in general, flux-calibrated since they are
measured with high-resolution ground-based spectrographs. We used corrected
PHOENIX model spectra in order to flux-calibrate the Ca II K SPI variations in
our sample (see Methods for details). We then converted the Ca II K fluxes to
disk-averaged powers by multiplying by $\pi R_*^2$.  

We show the flux-calibrated Ca II K residual spectra (top row) and summed
residual powers (rows 2 - 4) in Figure 1. We generate the residual spectra by
subtracting the average spectrum of the epoch from each individual spectrum.
The spectra are velocity-corrected to the rest frame of the observed. We then
calculate the residual powers by summing the residual spectrum across a 1.0
\AA\ bin centered on the Ca II K rest wavelength. The bottom row in Figure 1
shows the final SPI signals, i.e., residual powers phased to the planetary
orbital period.  The red lines are the best-fit sinusoid curves to the SPI
signal. We use the sinusoids to estimate the peak power in the signal and
orbital phase at which the peak power occurs (see Methods for further details).
SPI flux, power, and orbital phase offsets from the sub-planetary point, as
well as their 68\% confidence intervals, are given in Table 2.

\begin{figure*}[htbp]
   \centering
   \includegraphics[scale=.68,clip,trim=11mm 15mm 10mm 15mm,angle=0]{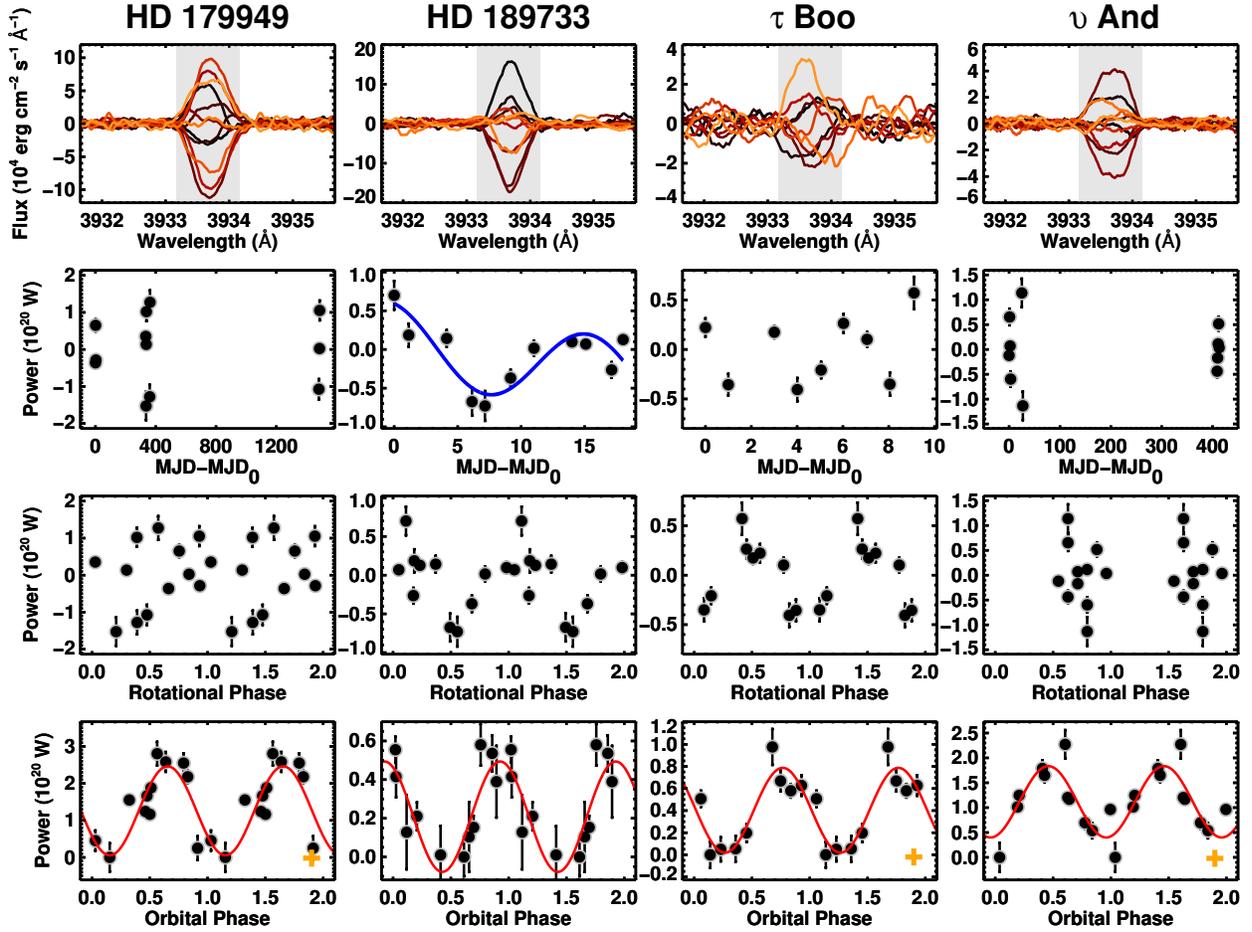}

   \caption{\fontfamily{phv}\selectfont{Ca II K residual spectra (top row),
summed residual power as a function of time (second row) and stellar rotational
phase (third row), and summed residual power as a function of planetary orbital
phase (bottom row).  The colors in the top row represent different nightly
observations: black is earlier in time and red is more recent (see
Supplementary Table 1). The residual spectra are smoothed by twenty pixels for
clarity and the line core flux integration range is shown in gray. The blue
curve in the rotational phase panel for HD 189733 is the best-fit sinusoid to
the rotational modulation$^{16}$. Note that $\tau$ Boo is synchronously
rotating with its planet. The best-fit sinusoids to the orbitally phased data
are shown in red. Objects marked with an orange cross do not have their
rotational modulation removed. Uncertainties are 1-$\sigma$ values and are
quadrature sums of the residual flux uncertainties. The power uncertainties
include the uncertainty of the stellar radius.}}

\end{figure*}

\bigskip

\noindent\textbf{Theories of magnetic SPI.} Magnetic SPI signals can be used to
estimate the strength of planetary magnetic fields by comparing the observed
power with predictions from magnetic SPI models.  Several analytical
predictions for the power released in magnetic SPI have been developed, the
first of which estimated the power by assuming that the available energy is
generated entirely in reconnection events between the planetary and stellar
magnetospheres$^2$. It was shown that the power (\textit{P}) at the boundary
between the stellar and planetary magnetospheres is 

\begin{equation}
P = \gamma \frac{\pi}{\mu} R_\text{p}^2 B_*^{4/3} B_\text{p0}^{2/3} v_\text{rel}
\end{equation}

\noindent where $R_\text{p}$ is the planetary radius, $B_*$ is the magnitude of
the stellar magnetic field at the orbital distance of the planet, $B_\text{p0}$
is the planetary surface magnetic field at the pole, $v_\text{rel}$ is the
relative velocity between the planet and the stellar magnetic field lines at
the planet's orbital distance, $\mu$ is the magnetic permeability of free
space, and $0 < \gamma < 1$ is an efficiency factor that depends on the
relative angle of the stellar and planetary magnetospheres$^{25,26}$. Assuming
$\gamma = 0.5$, it was demonstrated$^{26}$ that magnetic reconnection alone
produces powers $\approx 2-3$ orders of magnitude lower than the $\approx
10^{20}$ W estimate for the SPI signal observed around HD 179949$^{13}$. Even
if we assume $\gamma = 1$, $R_\text{p} = 1 R_\text{J}$, $B_* = 0.005$ G,
$B_\text{p0} = 100$ G, and $v_\text{rel} = 150$ km s$^{-1}$ we find a total
power of $\approx 2 \times 10^{18}$ W, two orders of magnitude lower than the
observed Ca II  K powers.  Planetary field strengths of order $\sim 10^4$ G are
needed to reach the observed powers. Because of this, we do not further
consider the reconnection scenario as a viable model.

Alfv\'{e}n wing SPI models have also been explored$^{5,32}$. It was found for
HD 179949 b that values of the planetary magnetic field strength $\approx
4000\times$ greater than Jupiter's magnetic field strength are needed to
reproduce the $10^{20}$ W estimated for the observed Ca II SPI signal$^{13}$.
A variety of Alfv\'{e}n wing scenarios were investigated for hot massive
planets, including the importance of the alignment between stellar and
planetary magnetospheres, and scaling laws were generated which estimated the
energy available in each case$^{5}$. For most cases, it was found that the
available power is $\approx 10^{19}$ W, slightly lower than needed to explain
the powers derived here, especially given the fact that the measured powers
include only that emitted by the Ca II  K line which is a fraction of the total
emitted power.

In another possible scenario, planets may increase the star's magnetic
helicity, which is a measure of the twisting of coronal field lines and is
generally conserved during the coronal field evolution putting a constraint on
the possible energy release process. Power estimates have been computed for
variations in coronal magnetic field helicity as induced by the passage of the
massive planet through the stellar magnetosphere$^{26}$. It was found that both
linear and non-linear force-free fields provide a similar amount of power
compared with the case of pure reconnection and thus are also inadequate to
explain the observations.  

The most promising scenario for producing the necessary power, which was
explored analytically with the goal of estimating how energy generated in
magnetic SPI can cause evaporation of planetary material, is the Poynting flux
across the base of a magnetic flux tube connecting the planetary surface to the
stellar surface$^{33}$. The power is generated by continuous stretching of the
magnetic footprint on the stellar surface due to the relative motion between
the planet and stellar magnetosphere.  It was shown that in this case the
available power is approximately

\begin{equation}
P \approx \frac{2 \pi}{\mu} f_{AP} R_\text{p}^2 B_\text{p0}^{2} v_\text{rel}
\end{equation}

\noindent where $R_p$, $B_\text{p0}$, $\mu$, and $v_\text{rel}$ have the same
meanings as in equation (1) and $f_{AP}$ is the fraction of the planetary
hemisphere covered by flux tubes$^{33}$. In this case, the total available
power is $\approx 10^{20} - 10^{21}$ W, sufficient to explain the powers
measured here. We consider this the most likely description of magnetic SPI for
hot giant planets as it is the only one that predicts the observed emitted
powers. We note that similar powers are obtained in numerical simulations of
hot giant planets involving similar magnetic interactions with their host
stars$^{28}$.

\begin{figure*}[tbh]
   \centering
   \includegraphics[scale=.70,clip,trim=13mm 60mm 10mm 85mm,angle=0]{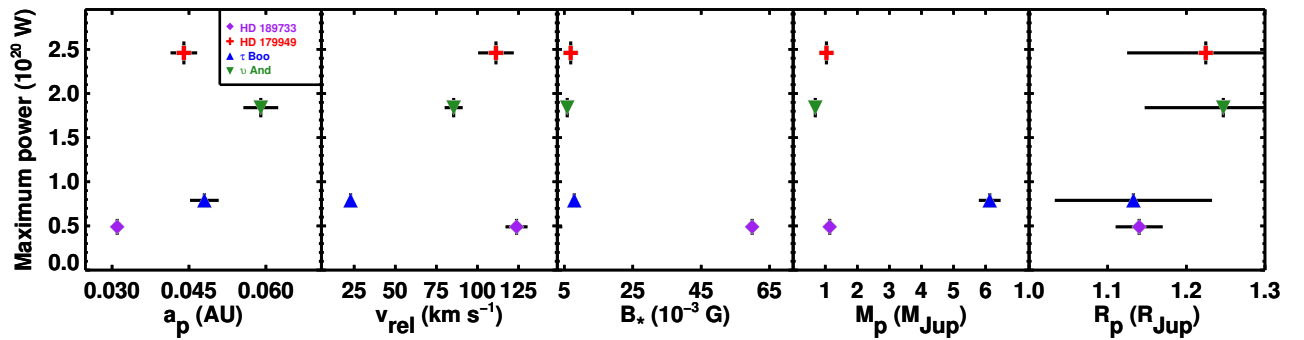}

   \caption{\fontfamily{phv}\selectfont{Observed powers in the Ca II  K line residuals as a function of
relevant magnetic SPI parameters. The powers range from $\approx 0.5$ to 2.2
$10^{20}$ W. Note that $B_*$ is the stellar magnetic field strength at the
orbital distance of the planet. Uncertainties for the system parameters are
1-$\sigma$ literature values. There are no obvious correlations between any of
the individual parameters and the measured powers.}}

\end{figure*}

Figure 2 shows the maximum Ca II K power as a function of relevant SPI
parameters from equation (2). There is no clear trend with any single system
parameter, although the two planets with the largest radii have the highest
powers. The range of observed powers only varies by a factor of $\approx 5$,
suggesting that the physical mechanism responsible for producing the Ca II  K
variations is roughly the same in all systems. If, for example, the
reconnection scenario given by equation (1) was producing the SPI signal in one
system and the flux tube scenario from equation (2) was active in another, we
might expect to see powers that differ by a factor of $\approx 100-1000$.

Another indicative feature of the observed magnetic SPI signals is the phase at
which the power is greatest: for all objects the peak power is reached between
$\phi_\text{max} \approx 0.5 - 0.9$, which corresponds to a phase lag between
the planet and the induced hot spot on the star of $\Delta\phi \approx
180^\circ$ and $30^\circ$.  These values are consistent with certain cases of
the linear and non-linear force-free flux tube models$^{26}$, although the
large phase lag observed for $\upsilon$ And is difficult to reproduce with
either model.  Regardless of the specific model interpretation, it is notable
that three of the four objects have phase lags within a fairly narrow range. 

\bigskip

\noindent\textbf{Planetary magnetic field strengths.}\ With the Ca II K powers, we
can calculate absolute field strengths for the planets using the favored flux
tube SPI model. All of the quantities in equation (2) are known or can be well
estimated and thus we can solve for $B_\text{p0}$ for each planet. In equation
(2), $f_\text{AP}$ depends on the ratio $B_*/B_\text{p0}$ and so equation (2) must
be solved numerically. For the field strengths we derive here, $f_\text{AP} =
0.02 - 0.20$. 

Since the observed Ca II  K power is a lower limit to the total power produced
by the SPI mechanism we need to estimate the fraction of the total SPI energy
radiated away in Ca II K. We accomplish this by comparing the SPI emission to
moderate solar flares, which are also powered by the release of magnetic
energy. Using the modeled dissipated energies for M-class solar flares$^{34}$,
which comprise $\approx 10\%$ of measured X-ray flares$^{35}$, and the energy
radiated in Ca II K for an M-class flare ($\approx 2.5 \times 10^{29}$
ergs$^{36}$), we estimate that Ca II K emits energy equal to $0.21\% \pm
0.08\%$ of the total energy dissipated in the flare, where the uncertainty is
the standard deviation of the M-class flare dissipated energies (assuming a
constant value for the emitted Ca II K energy). We calculate the planetary
field strengths within the $\pm 1\sigma$ boundaries for the Ca II K conversion
fraction.  We note that the fraction of optical emission line energy radiated
in Ca II K was found to be $\approx 14\%$ of the total energy observed during a
flare on HD 189733, which emitted a total of $>9 \times 10^{31}$ ergs$^{37}$.
The same measurement of Ca II K energy in the aforementioned M-class solar
flare, which emitted $\approx 2 \times 10^{30}$ ergs, is 13\%$^{36}$,
suggesting that the fraction of energy emitted by Ca II K is consistent across
this range of flare energies.

In Table 3 we report magnetic field strengths of the sample for different
values of $\epsilon$, the fraction of the total dissipated energy radiated away
in Ca II K. We give the $\epsilon = 100\%$ values as a lower-limit
reference. We derive the $1\sigma$ uncertainties by propagating the parameter
errors from Table 1 in quadrature. Note that, with the exception of $\upsilon$
And b, the uncertainties on individual values of $B_\text{p0}$ are similar to
the spread in $B_\text{p0}$ for different values of $\epsilon$. The measurement
uncertainty for $\upsilon$ And b is $\approx 90\%$ of the planetary field
strength value due to the large uncertainty in the stellar magnetic field.

We also include in Table 3 (last row) the surface polar planetary field
strengths calculated using the recent scaling relation for hot planets which
includes the extra heat from the central star$^{38}$. Building on previous work
which showed that the magnetic field strength of giant planets depends on their
internal heat flux$^{39,40}$, magnetic field strengths were estimated for hot
Jupiters which receive an incident flux $\geq 2 \times 10^8$ erg s$^{-1}$
cm$^{-2}$ and have $R_\text{p} \geq R_\text{J}$$^{38}$. This model included the
excess luminosity that results from heating by the central star$^{41}$. Since
the magnetic field scaling law relies on the internal heat flux of the
planet$^{39}$, the field strengths including heating from the central star are
$\approx 4-5\times$ larger than the nominal values which only consider the
cooling of the planet over time$^{40}$. The models which include the extra
heating assume for simplicity that the heat is deposited at the center of the
planet.  Off-center heat deposition may disrupt convective flows in the planet
interior or reduce the efficiency of conversion to magnetic energy, reducing
the magnetic field strength. Thus the extra heat values are likely upper
limits. On the other hand, it has been noted that the dynamo generation region
in giant planets may be close to the surface$^{30}$, which would allow more of
the deposited heat to contribute to the generation of a magnetic field.

\begin{figure*}[htbp]
   \centering
   \includegraphics[scale=1.0,clip,trim=10mm 20mm 0mm 55mm,angle=0]{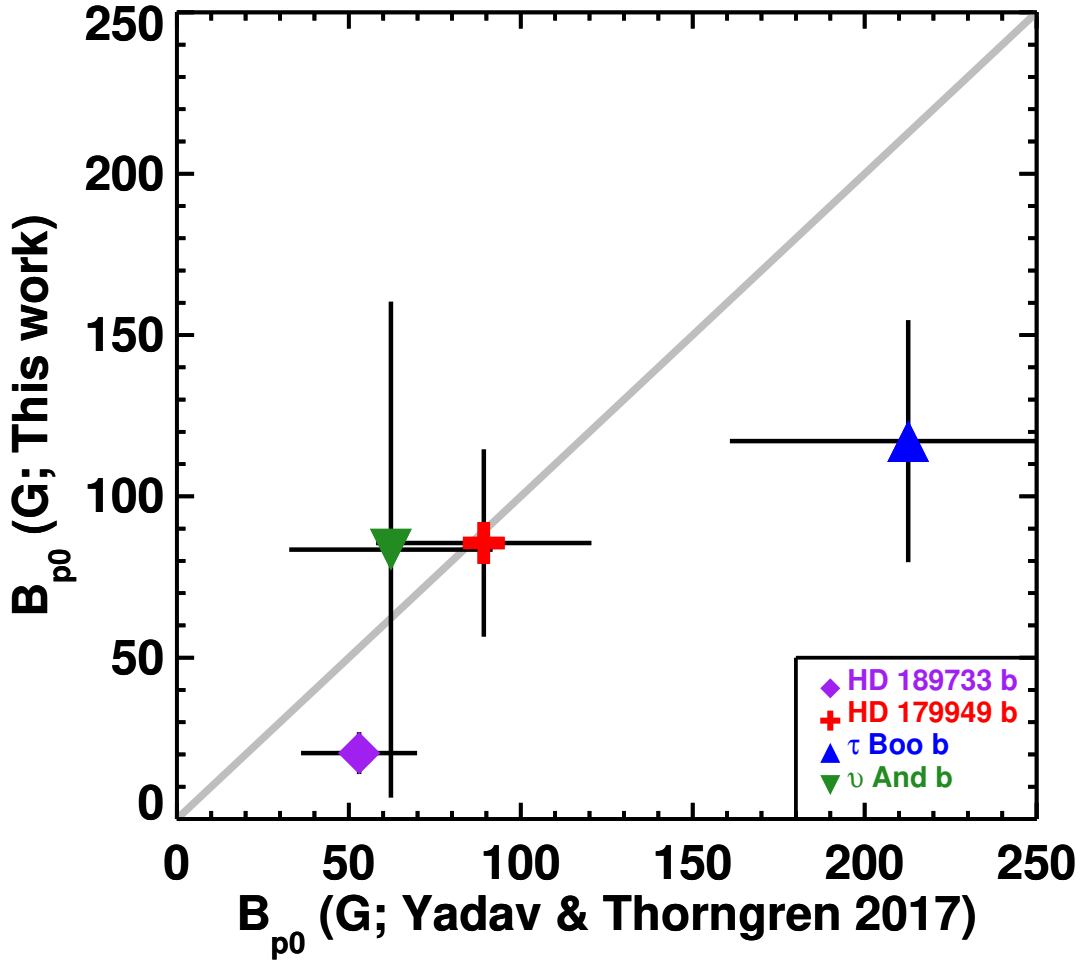}

   \caption{\fontfamily{phv}\selectfont{Magnetic field strengths from equation (2) and those calculated
using the extra heat deposition models$^{38}$ for the case of $\epsilon =
0.2\%$ from Table 3. The gray line is the line of equal values. Uncertainties
are 1-$\sigma$ values and are derived by propagating the parameter and power
errors in quadrature. There is some correspondence between the SPI field
strengths and those from the extra heat deposition models, supporting internal
heat flux descriptions of magnetic field generation in hot planets.}}

\end{figure*}

We plot the planetary field strengths from equation (2) against the planetary
field strengths calculated with the extra stellar heating models$^{38}$ in
Figure 3. There is some agreement between the SPI-derived fields and those from
the scaling relations which include the additional heat flux from the star.
More important is the fact that the values of $B_\text{p0}$ we derive here are
$\approx 2 - 8\times$ larger than predicted by rotation scaling and internal
heat flux evolution without considering the extra energy deposited into hot
planet interiors by their host stars$^{40}$. We suggest that the correspondence
shown here between the field strengths derived from the flux tube SPI scenario
and the extra heat deposition model supports the idea of hot massive planets
hosting much stronger magnetic fields than would be expected from only
considering the thermal evolution of the planet and their slow
rotation$^{30,31}$. More flux-calibrated SPI detections are needed for planets
with known radii (i.e., transiting planets) in order to enlarge the comparison
sample. 

Our results are in contrast with the small magnetic moments derived for hot
planets using models of exospheric Lyman-$\alpha$ absorption.  A magnetic
moment of $\approx 0.1 \mu_\text{Jup}$, where $\mu_\text{Jup}$ is the magnetic
moment of Jupiter, was found for HD 209458 b$^{42}$. The extra heat deposition
models estimate a polar surface magnetic field strength for HD 209458 b of
$\approx 49$ G$^{38}$, approximately three times stronger than Jupiter's
surface field.  Similarly, the magnetic moment for the hot Neptune GJ 436 b was
estimated to be $\approx 0.16 \mu_\text{Jup}$$^{43}$. These models, however,
rely on a number of highly uncertain parameters (e.g., planetary mass loss rate
and stellar wind density) and, in the case of HD 209458 b, a
low-signal-to-noise Lyman-$\alpha$ transmission spectrum. 

The large planetary field strengths we find for our sample have important
consequences for the detection of radio emission from the planets'
magnetospheres.  It has been suggested that the dense ionospheres of some hot
Jupiters may quench any radio waves generated by the electron-cyclotron maser
instability (ECMI) since the plasma frequency will be greater than that of the
radio emission$^{44,45}$. However, this is generally only the case for weak
planetary field strengths of the order $\approx 1 - 10$ G and planets with
extended ionospheres$^{46}$.  For the planetary field strengths we find here,
conditions are more favorable for the generation and escape of radio waves
produced by the ECMI.  The peak radio frequencies emitted by the ECMI for our
sample, which is $\nu_\text{peak} \approx 2.8 B_\text{p0}$ MHz$^3$, are 56,
240, 327, and 232 MHz for HD 189733 b, HD 179949 b, $\tau$ Boo b, and
$\upsilon$ And b, respectively. Radio observations of our sample near these
frequencies will be useful in confirming our derived field strengths which will
aid in illuminating the most dominant physical parameters for predicting giant
planet magnetic fields.

\bigskip

\noindent\textbf{Methods}\\

\noindent\textbf{Observations and data reduction.} High-resolution spectra ($R
\approx 65,000 - 110,000$) of our targets (Supplementary Table 1) were obtained
with NARVAL on the 2-meter T\'{e}lescope Bernard Lyot (TBL) or Gecko or
ESPaDOnS on the 3.6-meter Canada France Hawaii Telescope (CFHT). We give the
references for the original publications in Supplementary Table 1.  Data
reduction details can be found in these references but a brief summary is given
here. Standard reduction steps were performed for all exposures, including bias
subtraction, or dark subtraction in the case of the Gecko data, flat fielding,
and wavelength calibration using a reference lamp exposure. All data from
ESPaDOnS and NARVAL were reduced using the automated pipeline
\texttt{LIBRE-ESPRIT}. The typical signal-to-noise per pixel in the continuum
near Ca II K for an average nightly spectrum ranges from $\approx 300$ up to
$\approx 3000$. All spectra are corrected for the Earth's heliocentric motion,
the system's radial velocity, and the radial velocity induced by the planet on
the star.

All of the stellar and planetary parameters and their uncertainties are given
in Table 1. Uncertainties for the orbital periods and semi-major axes are
negligible and are omitted from the analysis. We estimate the radii of the
non-transiting planets using a probabilistic mass-radius relationship$^{56}$.
For hot Jupiters, the radius is also a function of incident stellar flux, with
more highly irradiated planets having larger radii for a given mass$^{41}$.
This results in a large radius spread of $\approx 15\%$ at a particular
mass$^{56}$ since most of the Jupiter-mass planets with known radii are hot
Jupiters. We thus consider the planetary radii to be uncertain at the level of
$15\%$.

\bigskip

\noindent\textbf{Calculating the Ca II K fluxes.} High-resolution spectra are
not, in general, flux calibrated, which makes it impossible to use the SPI
signals to derive planetary magnetic field strengths. Estimates of the surface
flux can be obtained by using the absolute fluxes from PHOENIX model spectra of
the same $T_\text{eff}$. These PHOENIX models have been used in a number of
previous studies for a similar purpose$^{57,58}$ and are currently the
gold-standard in stellar photosphere models for FGKM main-sequence stars.

In order to estimate the Ca II K surface fluxes of our objects, we used a grid
of PHOENIX model spectra$^{59}$ with log$g = 4.5$ and $\text{[Fe/H]}=0.0$.
Before applying the PHOENIX spectra to the data, it is important to understand
how well the model surface fluxes predict measured flux values.

To accomplish this we retrieved the latest version of the Next Generation
Spectral Library (NGSL) from the MAST
database\footnote{https://archive.stsci.edu/prepds/stisngsl/}.  The NGSL
contains 374 stars with spectra covering $\lambda = 2000 - 10000$ \AA\ at $R
\approx 1000$, observed using the Space Telescope Imaging Spectrograph onboard
the \textit{Hubble Space Telescope}. To ensure that we only used NGSL stars
with well-determined parameters, we cross-referenced the NGSL objects with
those from the PASTEL catalog$^{60}$, resulting in 295 matches.
Since our targets are comprised of main sequence FGK stars, we only selected
objects from the remaining sample with 4500 \text{K} $< T_\text{eff} < 7000$ K,
$4.1 < \text{log}g < 4.7$, and $-1.0 < \text{[Fe/H]} < 1.0$. These cuts
resulted in 38 final objects.

Converting observed fluxes to surface fluxes requires knowledge of the object's
distance and radius. All of the final 38 objects have well known distances, as
they all reside $d_* < 100$ pc from the sun. In order to derive radii for the
sample, we used the $T_\text{eff}$ and log$g$ values from the PASTEL catalog
and a recent $T_\text{eff} - M_*$ relationship$^{61}$. We then solved
for $R_*$ using the measured log$g$ values and the interpolated $M_*$ values.

Once the NGSL spectra were converted into surface flux, we took the mean flux
value in two continuum bands between 3885-3915 and 3980-4010 \AA. We performed
the same steps to the PHOENIX model spectra, which we convolved down to the
spectral resolution of the NGSL spectra. An example of a comparison between the
NGSL star HD 21742 with $T_\text{eff} = 5160$ K and the corresponding PHOENIX
model of the same $T_\text{eff}$ is shown in Supplementary Figure 1.  The
resulting mean NGSL and PHOENIX model fluxes are shown in the top panel of
Supplementary Figure 2; the bottom panel shows the ratio of the model fluxes to
the NGSL fluxes.

Overall, the measured fluxes are well approximated by the PHOENIX models.
However, there is a trend in the flux ratio: the models tend to over-predict
the measured values for $T_\text{eff} \lesssim 5500$ K by $\approx 10 - 30\%$
and only slightly over-predict them for larger $T_\text{eff}$. A typical
example of the over-predicted flux is shown in Supplementary Figure 1.  To
account for this trend in $T_\text{eff}$ we fit a power law to the
$F_\text{mod}/F_\text{CaII}$ values in the lower panel. The fit is performed
using an MCMC procedure based on affine-invariant sampling$^{62,63}$.  The
best-fit result and 68\% confidence intervals are shown with the solid orange
line and gray band, respectively.  This fit is applied to all of the Ca II
flux derivations for our SPI sample. We note that for most objects the
correction is $< 15\%$.

The exact cause of the larger model discrepancies at lower $T_\text{eff}$ is
unknown.  The NGSL spectra were observed with the $52''\times0.2''$ slit and
there is no evidence of a systematic error in the flux calibration as a
function of stellar color. PHOENIX model surface fluxes between 3900 \AA\ and
4000 \AA\ for log$g = 4.0 - 5.0$ only differ by at most 15\%, ruling out the
constant log$g$ of the models as the cause. The difference is likely due to the
LTE treatment of most atomic species in the models, which can significantly
change the flux in the wings of strong lines, such as Ca II K.

\bigskip 

\noindent\textbf{Emitted power estimates.} All of the published spectra we use
were continuum normalized. In order to flux-calibrate a spectrum, a flux
correction factor is needed that is a function of wavelength across the desired
range. Our approach was as follows:

\begin{enumerate}

\item Interpolate between PHOENIX spectra to produce a model spectrum with approximately
the same $T_\text{eff}$ as the object.

\item Perform a linear fit to the continuum surrounding the Ca II  H and K lines of
the model spectrum.

\item Multiply the normalized object spectrum by the linear flux correction
factor.

\item Multiply the flux-calibrated object spectrum by the power law adjustment
factor.

\item Subtract the mean flux-calibrated spectrum of the epoch from all
individual spectra from the same epoch.

\item Sum the residual core flux across a 1 \AA\ band centered on the rest
wavelength of the line.

\item Multiply the core flux by $\pi R_*^2$ to find the power averaged over the
observable stellar disk.

\end{enumerate}

We show an example of steps 1-3 in the flux calibration process in
Supplementary Figure 3.  The Gecko spectra for HD 179949 and $\upsilon$ And
cover a subset of the Ca II  H and K order of ESPaDOnS and NARVAL spectra and
cannot be normalized in the same way using step 2. We made minor continuum
corrections to the Gecko orders by multiplying the continuum by the ratio
between the Gecko spectrum and an average spectrum of the same object taken
with ESPaDOnS from a separate epoch. We then applied the flux calibration
vector from the ESPaDOnS epochs to the Gecko spectra, which then have the same
stellar continuum shape as the ESPaDOnS spectra.

We note that the scatter of the observed fluxes compared with the power law fit
in Supplementary Figure 2 is larger ($\approx 15\%$) than the spread in surface
flux of models within the $T_\text{eff}$ uncertainty range ($\approx \pm 50$
K). Thus we do not consider the linear continuum fit or $T_\text{eff}$ as
sources of uncertainty in the surface fluxes.

We generated the residual spectra by subtracting the flux-calibrated average
spectrum for the epoch from each individual spectrum from the same epoch.  For
example, there are three separate observing epochs for HD 179979. Any small
offsets from zero, which are typically $5-10\%$ of the residual flux level, in
the continuum are then removed with a low order polynomial. Note that we only
calculated residual fluxes for the Ca II  K line due to the higher
signal-to-noise for some objects and the strong correlation between Ca II H and
K residual measurements$^{12}$. Once the spectra are corrected, the residual Ca
II K fluxes are summed across a 1.0 \AA\ wide band centered on 3933.66 \AA. We
then converted the fluxes to powers by multiplying by $\pi R_*^2$.

We calculated the uncertainties on the power estimates by combining in
quadrature the $1\sigma$ flux error from the residual spectra as measured
outside of the line core with the uncertainty in the stellar radius and a 15\%
uncertainty in the model flux correction factor.  The correction factor
uncertainty is the standard deviation of the residuals to the power law fit in
Supplementary Figure 2.

We subtracted rotational modulation attributed to the star from the residuals for
HD 189733$^{16}$. HD 179949 and $\upsilon$ And do not show evidence of
rotational modulation in the epochs chosen for this analysis.  $\tau$ Boo b has
an orbital period similar to its star's rotational period.  However,
photometric variations observed for $\tau$ Boo can more plausibly be attributed
to the planet rather than stellar rotation due to their persistence across
multiple years and consistency with the planetary orbital period and
phase$^{17}$. Based on this evidence it is most likely that the Ca II K
variations are due to the planet.

We fit a sinusoid to the orbitally phased signals in order to derive estimates
of the maximum power and phase offset. Although SPI variations may not be fully
described by a simple sinusoid, we chose this functional form to avoid
over-fitting the small number of data points. The period was fixed to the
planet's orbital period, leaving the phase offset and amplitude as the free
parameters.  The fits were performed with the same MCMC procedure used to derive
the flux correction factor. We chose the peak in the model sine curve and its
phase offset, along with each parameter's 68\% confidence intervals, as the
maximum power and phase offset for the observed SPI signals. The mean power was
taken directly from the data where the uncertainty is simply the standard
deviation of the mean. 

\bigskip

\noindent\textbf{Data Availability.} The data that support the plots within this
paper and other findings of this study are available from the corresponding
author upon reasonable request. The reduced spectra used here are also publicly
available via the PolarBase archive and the CFHT data archive.

\bigskip

\noindent\textbf{References}\\
\scriptsize
\begin{enumerate}

\item Wood, B. E., M\"{u}ller, H R, Zank, G. P, Linsky, J. L, Redfield, S.
New Mass-Loss Measurements from Astrospheric Ly$\alpha$ Absorption.
\textit{Astrophys. J.} \textbf{628,} L143-L146 (2005)

\item Cuntz, M., Saar, S., \& Zdzislaw, E. On Stellar Activity Enhancement Due
to Interactions with Extrasolar Giant Planets. \textit{Astrophys. J.}
\textbf{533,} L151-L154 (2000)

\item Stevens, I. R. Magnetospheric radio emission from extrasolar giant
planets: the role of the host stars. \textit{Mon. Not. R. Astron. Soc.}
\textbf{356,} 1053-1063 (2005) 

\item Vidotto, A. A., Jardine, M., \& Helling Ch. Transit variability in bow
shock-hosting planets. \textit{Mon. Not. R. Astron. Soc.} \textbf{414,}
1573-1582 (2011)

\item Strugarek, A. Assessing Magnetic Torgues and Energy Fluxes in Close-in
Star-Planet systems. \textit{Astrophys. J.} \textbf{833,} 140-152 (2016)

\item Rogers, T. M., \& McElwaine, J. N. The Hottest Hot Jupiters May Host
Atmospheric Dynamos. \textit{Astrophys. J. Lett.} \textbf{841,} L26-L32 (2017) 

\item Cohen, O., et al. The Interaction of Venus-like, M-dwarf Planets with the
Stellar Wind of Their Host Star. \textit{Astrophys. J.} \textbf{806,} 41-51 (2015) 

\item Jakosky, B. M., et al. Mars' atmospheric history derived from
upper-atmosphere measurements of $^{38}$Ar/$^{36}$Ar. \textit{Science}
\textbf{355,} 1408-1410 (2017) 

\item Blackman, E. G., \& Tarduno, J. A. Mass, energy, and momentum capture
from stellar winds by magnetized and unmagnetized planets: implications for
atmospheric erosion and habitability. \textit{Mon. Not. R. Astron. Soc.}
\textbf{481,} 5146-5155 (2018)

\item Lazio, T. J. W., et al. 2016, Planetary Magnetic Fields: Planetary interiors and habitability, 
final report prepared by the Keck Institute of Space Studies

\item Shkolnik, E. L., \& Llama, J. \textit{Handbook of Exoplanets} Signatures of Star-Planet Interactions (Springer, 2017)

\item Shkolnik, E., Walker, G. A. H., \& Bohlender, D. A. Evidence for Planet-induced 
Chromospheric Activity on HD 179949. \textit{Astrophys. J.} \textbf{597,} 1092-1096 (2003)

\item Shkolnik, E., Walker, G. A. H., Bohlender, D. A., Gu, P.-G., \& K\"{u}rster, M. Hot Jupiters and Hot Spots:
The Short- and Long-Term Chromospheric Activity on Stars with Giant Planets. \textit{Astrophys. J.} \textbf{622,} 
1075-1090 (2005)

\item Gurdemir, L., Redfield, S., \& Cuntz, M. Planet-Induced Emission Enhancements in HD 179949: Results from McDonald
Observations. \textit{Publ. Astron. Soc. Pac.} \textbf{29,} 141-149 (2012)

\item Shkolnik, E., Bohlender, D. A., Walker, G. A. H., \& Collier Cameron, A. The On/Off Nature of 
Star-Planet Interactions. \textit{Astrophys. J.} \textbf{676,} 628-638 (2008)

\item Cauley, P. W., Shkolnik, E. S., Llama, J., Bourrier, V., \& Moutou, C. Evidence of Magnetic 
Star-Planet Interactions in the HD 189733 System from Orbitally Phased Ca II K Variations. 
\textit{Astron. J.} \textbf{156,} 262-273 (2018)

\item Walker, G. A. H., et al. MOST Detects Variability on $\tau$ Bootis A Possibly Induced by 
its Planetary Companion. \textit{Astron. Astrophys.} \textbf{482,} 691-697 (2008)

\item Pagano, I., et al. CoRoT-2a Magnetic Activity: Hints for Possible Star-Planet Interaction. 
\textit{Earth Moon and Planets} \textbf{105,} 373-378 (2009)

\item Scandariato, G., et al. A Coordinated Optical and X-ray Spectroscopic Campaign on HD 179949: Searching for
Planet-Induced Chromospheric and Coronal Activity. \textit{Astron. Astrophys.} \textbf{552,} 7-20 (2013)

\item Maggio, A., et al. Coordinated X-Ray and Optical Observations of Star-Planet Interaction 
in HD 17156. \textit{Astrophy. J.} \textbf{811,} L2-L7 (2015)

\item Pillitteri, I., et al. FUV Variability of HD 189733. Is the Star Accreting Material 
from its Hot Jupiter? \textit{Astrophys. J.} \textbf{805,} 52-70 (2015)

\item Cranmer, S. R., \& Saar, S. H. Exoplanet-Induced Chromospheric Activity: Realistic 
Light Curves from Solar-type Magnetic Fields. Preprint at https://arxiv.org/pdf/astro-ph/0702530.pdf (2007)

\item Llama, J., et al. Exoplanet Transit Variability: Bow Shocks and Winds Around HD 189733 b. 
\textit{Mon. Not. R. Astron. Soc.} \textbf{436,} 2179-2187 (2013)

\item Fares, R., et al. A Small Survey of the Magnetic Fields of Planet-Host Stars. 
\textit{Mon. Not. R. Astron. Soc.} \textbf{435,} 1451-1462 (2013)

\item Lanza, A. F. Stellar Coronal Magnetic Fields and Star-Planet Interaction. 
\textit{Astron. Astrophys.} \textbf{505,} 339-350 (2009)

\item Lanza, A. F. Star-Planet Magnetic Interaction and Activity in Late-Type Stars 
with Close-In Planets. \textit{Astron. Astrophys.} \textbf{544,} 23-39 (2012)

\item Cohen, O., et al. The Dynamics of Stellar Coronae Harboring Hot Jupiters. 
I. A Time-Dependent Magnetohydrodynamic Simulations of the Interplanetary Environment in 
the HD 189733 Planetary System. \textit{Astrophys. J.} \textbf{733,} 67-79 (2011)

\item Scharf, C. A. Possible Constraints on Exoplanet Magnetic Field Strengths from Planet-Star
Interaction. \textit{Astrophys. J.} \textbf{722,} 1547-1555 (2010)

\item van Haarlem, M. P., et al. LOFAR: The LOw-Frequency ARray. \textit{Astron. Astrophys.}
\textbf{556,} 2-55 (2013)

\item Zaghoo, M., \& Collins, G. W. Size and Strength of Self-Excited Dynamos in Jupiter-like
Extrasolar Planets. \textit{Astrophys. J.} \textbf{862,} 19-29 (2018)

\item S\'{a}nchez-Lavega, A. The MAgnetic Field in Giant Extrasolar Planets. \textit{Astrophys. J.}
\textbf{609,} L87-L90 (2004)

\item Saur, J., Grambusch, T., Duling, S., Neubauer, F. M., \& Simon, S. Magnetic Energy
Fluxes in Sub-Alfv\'{e}nic Planet-Star and Moon-Planet Interactions. \textit{Astron. Astrophys.}
\textbf{552,} 119-139 (2013)

\item Lanza, A. F. Star-Planet Magnetic Interactions and Evaporation of Planetary Atmospheres.
\textit{Astron. Astrophys.} \textbf{557,} 31-44 (2013)

\item Aschwanden, M. J., Xu, Y., \& Jing, J. Global Energetics of Solar Flares. I. Magnetic
Energies. \textit{Astrophys. J.} \textbf{797,} 50-85 (2014)

\item Veronig, A., Temmer, M., Hanslmeier, A., Otruba, W., \& Messerotti, M. Temporal Aspects
and Frequency Distributions of Solar Soft X-ray Flares. \textit{Astron. Astrophys.}
\textbf{382,} 1070-1080 (2002)

\item Johns-Krull, C. M., Hawley, S. L., Basri, G., \& Valenti, J. A. Hamilton Echelle Spectroscopy
of the 1993 March 6 Solar Flare. \textit{Astrophys. J. Supp.} \textbf{112,} 221-243 (1997)

\item Klocov\'{a}, T., Czesla, S., Khalafinejad, S., Wolter, U., \& Schmitt, J. H. M. M.
Time-Resolved UVES Observations of a Stellar Flare on the Planet Host HD 189733 During
Primary Transit. \textit{Astron. Astrophys.} \textbf{607,} 66-78 (2017)

\item Yadav, R. K., \& Thorngren, D. P. Estimating the Magnetic Field Strengths in Hot
Jupiters. \textit{Astrophys. J. Lett.} \textbf{849,} L12-L16 (2017)

\item Christensen, U. R., Holzwarth, V., \& Reiners, A. Energy Flux Determines Magnetic Field
Strength of Planets and Stars. \textit{Nature} \textbf{457,} 167-169 (2009)

\item Reiners, A., \& Christensen, U. R. A Magnetic Field Evolution Scenario for Brown
Dwarfs and Giant Planets. \textit{Astron. Astrophys.} \textbf{522,} 13-20 (2010)

\item Thorngren, D. P., \& Fortney, J. J. Bayesian Analysis of Hot-Jupiter Radius
Anomalies: Evidence for Ohmic Dissipation? \textit{Astron. J.} \textbf{155,} 214-224 (2018)

\item Kislyakova, K. G., Holmstr\"{o}m, M., Lammer, H., Odert, P., \& Khodachenko, M. L.
Magnetic Moment and Plasma Environment of HD 209458b as Determined from Ly$\alpha$
Observations. \textit{Science} \textbf{346,} 981-984 (2014)

\item Bourrier, V., Lecavelier des Etangs, A., Ehrenreich, D., Tanaka, Y. A., \& Vidotto, A. A.
An Evaporating Planet in the Wind: Stellar Wind Interactions with the Radiatively Braked
Exosphere of GJ 436 b. \textit{Astron. Astrophys.} \textbf{591,} 121-135 (2016)

\item Weber, C., et al. How Expanded Ionospheres of Hot Jupiters Can Prevent Escape
of Radio Emission Generated by the Cyclotron Maser Instability. \textit{Mon. Not. R. Astron. Soc.}
\textbf{469,} 3505-3517 (2017)

\item Daley-Yates, S., \& Stevens, I. R. Inhibition of the Electron Cyclotron Maser
Instability in the Dense Magnetosphere of a Hot Jupiter. \textit{Mon. Not. R. Astron. Soc.}
\textbf{479,} 1194-1209 (2018)

\item Weber, C., et al. Supermassive Hot Jupiters Provide More Favourable Conditions for
the Generation of Radio Emission via the Cyclotron Maser Instability - A Case Study Based
on Tau Bootis b. \textit{Mon. Not. R. Astron. Soc.} \textbf{480,} 3680-3688 (2018)

\item Butler, R. P., et al. Catalog of Nearby Exoplanets. \textit{Astrophys. J.} \textbf{646,} 505-522 (2006)

\item Fares, R., et al. Magnetic Field, Differential Rotation and Activity of the Hot-Jupiter-Hosting
Star HD 179949. \textit{Mon. Not. R. Astron. Soc.} \textbf{423,} 1006-1017 (2012)

\item Fares, R., et al. MOVES - I. The Evolving Magnetic Field of the Planet-Hosting
Star HD 189733. \textit{Mon. Not. R. Astron. Soc.} \textbf{471,} 1246-1257 (2017)

\item Bouchy, F., et al. ELODIE Metallicity-Biased Search for Transiting Hot Jupiters. II. A Very Hot
Jupiter Transiting the Bright K Star HD 189733. \textit{Astron. Astrophys.} \textbf{444,} L15-L19 (2005)

\item Winn, J., et al. The Transit Light Curve Project. V. System Parameters and Stellar Rotation
Period of HD 189733. \textit{Astron. J.} \textbf{133,} 1828-1835 (2007)

\item Boisse, I., et al. Stellar Activity of Planetary Host Star HD 189733. \textit{Astron. Astrophys.}
\textbf{495,} 959-966 (2009)

\item Catala, C., Donati, J.-F., Shkolnik, E. S., Bohlender, D., \& Alecian, E. The Magnetic
Field of the Planet-Hosting Star $\tau$ Bootis. \textit{Mon. Not. R. Astron. Soc.} \textbf{374,}
L42-L46 (2007)

\item Jeffers, S. V., et al. The Relation Between Stellar Magnetic Field Geometry and
Chromospheric Activity Cycles - II. The Rapid 120-Day Magnetic Cycle of $\tau$ Bootis.
\textit{Mon. Not. R. Astron. Soc.} \textbf{479,} 5266-5271 (2018)

\item Marsden, S. C., et al. A BCool Magnetic Snapshot Survey of Solar-Type Stars.
\textit{Mon. Not. R. Astron. Soc.} \textbf{444,} 3517-3536 (2014)

\item Chen, J., \& Kipping, D. Probabilistic Forecasting of the Masses and Radii of Other
Worlds. \textit{Astrophys. J.} \textbf{834,} 17-30 (2017)

\item Mittag, M., Schmitt, J. H. M. M., \& Schr\"{o}der, K.-P. Ca II H+K Fluxes from S-Indices
of Large Samples: A Reliable and Consistent Conversion Based on PHOENIX Model Atmospheres.
\textit{Astron. Astrophys.} \textbf{549,} 117-129 (2013)

\item Scandariato, G., et al. HADES RV Programme with HARPS-N at TNG. IV. Time Resolved Analysis of
the Ca II H\&K and H$\alpha$ Chromospheric Emission of Low-Activity Early-Type M Dwarfs.
\textit{Astron. Astrophys.} \textbf{598,} 28-42 (2017)

\item Husser, T.-O., et al. A New Extensive Library of PHOENIX Stellar Atmospheres and Synthetic
Spectra. \textit{Astron. Astrophys.} \textbf{553,} 6-15 (2013)

\item Soubiran, C., Le Campion, J.-F., Brouillet, N., \& Chemin, L. The PASTEL Catalogue:
 2016 Version. \textit{Astron. Astrophys.} \textbf{591,} 118-125 (2016)

\item Eker, Z., et al. Main-Sequence Effective Temperatures from a Revised Mass-Luminosity
Relation Based on Accurate Properties. \textit{Astron. J.} \textbf{149,} 131-147 (2015)

\item Goodman, J., \& Weare, J. Ensemple Samplers with Affine Invariance.
\textit{Comm. App. Math. Comp.} \textbf{5,} 65-80 (2010)

\item Foreman-Mackey, D., Hogg, D. W., Lang, D., \& Goodman, J. emcee: The MCMC Hammer.
Preprint at https://arxiv.org/abs/1202.3665 (2012)

\end{enumerate}

\bigskip

\noindent\textbf{Acknowledgments.} We thank Travis Barman for insightful
discussions concerning details of the PHOENIX models. P.W.C. and E.L.S.
acknowledge support from NASA Origins of the Solar System grant No. NNX13AH79G
(P. I. Shkolnik).  This work has made use of NASA's Astrophysics Data System
and used the facilities of the Canadian Astronomy Data Centre operated by the
National Research Council of Canada with the support of the Canadian Space
Agency.

\bigskip

\noindent\textbf{Author Contributions.} This work made use of archived data. E.
L. S. was responsible for most of the original observing proposals and data
collection.  P. W. C. was responsible for the flux calibration and SPI signal
analysis, as well as the manuscript preparation. J. L. was responsible for some
original SPI signal analysis and also contributed to the manuscript. A. F. L.
provided interpretation of the SPI theories and oversight of the theory
application.  All authors contributed material to the manuscript.

\bigskip

\noindent\textbf{Competing interests.} The authors declare no competing interests.

\clearpage

\begin{table*}
\centering
\scriptsize
\fontfamily{phv}\selectfont{ \begin{tabular}{lccccccccc}

 &$T_\text{eff}$&$R_*$&$P_\text{rot}$&$B_*(r=R_*)$&$M_\text{pl}$&$R_\text{p}$&$P_\text{orb}$&$a_\text{p}$& \\
Object&(K)&($R_\odot$)&(days)&(G)&($M_\text{Jup}$)&($R_\text{Jup}$)&(days)&(AU)&References \\
\hline
HD 179949 & $6170 \pm 50$ & $1.23 \pm 0.03$ & $11.0 \pm 0.8$ & $3.2 \pm 0.3$ & $1.04 \pm 0.08$ & $1.22 \pm 0.18$ & 3.0925 & 0.044 & 47,48\\
HD 189733 & $5040 \pm 50$ & $0.76 \pm 0.01$ & $11.9 \pm 0.16$ & $27.0 \pm 3.0$ & $1.14 \pm 0.03$ & $1.14 \pm 0.03$ & 2.2186 & 0.031 & 49,50,51,52\\
$\tau$ Boo & $6387 \pm 50$ & $1.48 \pm 0.08$ & $3.7 \pm 0.1$ & $2.6 \pm 0.2$ & $6.13 \pm 0.34$ & $1.13 \pm 0.17$ & 3.3124 & 0.048 & 47,53,54\\
$\upsilon$ And & $6213 \pm 50$ & $1.64 \pm 0.04$ & $12.0 \pm 0.1$ & $2.5 \pm 1.1$ & $0.69 \pm 0.06$ & $1.25 \pm 0.19$ & 4.6170 & 0.059 & 47,55 \\

\end{tabular}}

\caption{\fontfamily{phv}\selectfont{ \scriptsize \textbf{SPI system
parameters.} All uncertainties are 1-$\sigma$ values. The masses for HD 179949
b and $\tau$ Boo b were derived from $M_p$sin$i$, stellar rotational periods,
$v$sin$i$, and orbital inclination assuming alignment with the stellar spin
axis. The radii for HD 179949, $\tau$ Boo b, and $\upsilon$ And b were
estimated using a probabilistic mass-radius relationship$^{56}$. Significant
differential rotation has been noted for $\tau$ Boo$^{53}$ and HD
179949$^{48}$. Here we take the rotation period for the polar latitudes in both
systems.}}

\end{table*}

\begin{table}
\centering
\scriptsize
\fontfamily{phv}\selectfont{ \begin{tabular}{lccccc}

&mean($F_\text{K}$)&max($F_\text{K}$)&mean($P_\text{K}$)&max($P_\text{K}$)& \\
Object&(10$^4$ erg cm$^{-2}$ s$^{-1}$)&(10$^4$ erg cm$^{-2}$ s$^{-1}$)&
(10$^{20}$ W)&(10$^{20}$ W)&$\phi_\text{max}$\\
\hline
HD 179949 & $6.67 \pm 1.18$ & $10.76_{-0.28}^{+0.28}$ & $1.53 \pm 0.27$ & $2.46_{-0.13}^{+0.13}$ & $0.65_{-0.01}^{+0.01}$\\
HD 189733 & $3.22 \pm 0.77$ & $5.76_{-0.63}^{+0.62}$ & $0.28 \pm 0.07$ & $0.49_{-0.09}^{+0.09}$ & $0.92_{-0.02}^{+0.02}$\\
$\tau$ Boo & $1.25 \pm 0.35$ & $2.43_{-0.12}^{+0.12}$ & $0.41 \pm 0.11$ & $0.79_{-0.08}^{+0.08}$ & $0.77_{-0.02}^{+0.02}$ \\
$\upsilon$ And & $2.64 \pm 0.44$ & $4.25_{-0.10}^{+0.10}$ & $1.14 \pm 0.19$ & $1.84_{-0.11}^{+0.11}$ & $0.46_{-0.01}^{+0.01}$\\
\end{tabular}}

\caption{\scriptsize \fontfamily{phv}\selectfont{ \textbf{Ca II K absolute
fluxes ($F_K$) and powers ($P_K$).} The uncertainties on the mean flux and mean
power are the standard deviation of the mean for each phased SPI signal. The
uncertainty for the maximum flux is the 1-$\sigma$ error for the individual
value. Uncertainties on the maximum power and peak phase are derived from the
MCMC sinusoid fitting (see Methods). The sub-planetary point is taken to be
where $\phi = 0$.}} 

\end{table}

\begin{table}
\centering
\scriptsize
\fontfamily{phv}\selectfont{ \begin{tabular}{ccccc}

$\epsilon\ = E_\text{CaIIK}/E_\text{tot}$&HD 189733 b&HD 179949 b&$\tau$ Boo b&$\upsilon$ And b\\
\hline
100\% & $0.4 \pm 0.1$ & $1.9 \pm 0.7$ & $2.7 \pm 0.9$ & $1.9 \pm 1.8$ \\
0.12\% & $27 \pm 8$ & $111 \pm 38$ & $163 \pm 52$ & $118 \pm 109$\\
0.20\% & $20 \pm 7$ & $86 \pm 29$ & $117 \pm 38$ & $83 \pm 77$\\
0.28\% & $17 \pm 5$ & $68 \pm 23$ & $95 \pm 30$ & $68 \pm 63$\\
 & & & & \\
Yadav \& Thorngren (2017) & $53 \pm 17$ & $89 \pm 31$ & $213 \pm 52$ & $62 \pm 30$\\
\end{tabular}}

\caption{\scriptsize \fontfamily{phv}\selectfont{ \textbf{Surface polar
planetary magnetic field ($B_\text{p0}$) strengths.} Magnetic fields are
calculated using the flux tube model from equation (2) and the extra heat
deposition theory$^{38}$. Field strengths are given in units of Gauss.  All
uncertainties are 1-$\sigma$ values and are derived by combining the various
parameter uncertainties from equation (2) in quadrature.}}

\end{table}

\clearpage

\renewcommand{\figurename}{\fontfamily{phv}\selectfont{Supplementary Figure}}
\renewcommand{\tablename}{\fontfamily{phv}\selectfont{Supplementary Table}}
\setcounter{figure}{0}
\setcounter{table}{0}

\noindent\textbf{Supplementary tables and figures.} Supplementary information
including a log of the observations (Supplementary Table 1) and graphics
showing the flux calibration method (Supplementary Figures 1 - 3).

\begin{table}[h!]
\centering
\scriptsize
\fontfamily{phv}\selectfont{ \begin{tabular}{lccccl}

Object&Observation Date&Instrument&Telescope&$\lambda/\Delta\lambda$&Reference \\

\hline

HD 179949 & 2001 Aug 25 &  Gecko & CFHT & 110,000 & 12 \\
 & 2001 Aug 26 &  Gecko & CFHT & 110,000 &  \\
 & 2001 Aug 28 &  Gecko & CFHT & 110,000 &  \\
 & 2002 Jul 25 &  Gecko & CFHT & 110,000 &  \\
 & 2002 Jul 28 &  Gecko & CFHT & 110,000 &   \\
 & 2002 Jul 29 &  Gecko & CFHT & 110,000 &   \\
 & 2002 Aug 20 &  Gecko & CFHT & 110,000 &   \\
 & 2002 Aug 22 &  Gecko & CFHT & 110,000 &   \\
 & 2005 Sep 16 &  ESPaDOnS & CFHT & 80,000 & 15 \\
 & 2005 Sep 20 &  ESPaDOnS & CFHT & 80,000 &   \\
 & 2005 Sep 21 &  ESPaDOnS & CFHT & 80,000 &   \\
HD 189733 & 2013 Aug 4 & NARVAL & TBL & 65,000 & 16,48\\
& 2013 Aug 5 & NARVAL & TBL & 65,000 &   \\
& 2013 Aug 8 & NARVAL & TBL & 65,000 &   \\
& 2013 Aug 10 & NARVAL & TBL & 65,000 &   \\
& 2013 Aug 11 & NARVAL & TBL & 65,000 &   \\
& 2013 Aug 13 & NARVAL & TBL & 65,000 &   \\
& 2013 Aug 15 & NARVAL & TBL & 65,000 &   \\
& 2013 Aug 18 & NARVAL & TBL & 65,000 &   \\
& 2013 Aug 19 & NARVAL & TBL & 65,000 &   \\
& 2013 Aug 21 & NARVAL & TBL & 65,000 &   \\
& 2013 Aug 22 & NARVAL & TBL & 65,000 &   \\
$\tau$ Boo & 2006 Jun 9 & ESPaDOnS & CFHT & 80,000 & 15 \\
& 2006 Jun 10 & ESPaDOnS & CFHT & 80,000 &   \\
& 2006 Jun 12 & ESPaDOnS & CFHT & 80,000 &   \\
& 2006 Jun 13 & ESPaDOnS & CFHT & 80,000 &   \\
& 2006 Jun 14 & ESPaDOnS & CFHT & 80,000 &   \\
& 2006 Jun 15 & ESPaDOnS & CFHT & 80,000 &   \\
& 2006 Jun 16 & ESPaDOnS & CFHT & 80,000 &   \\
& 2006 Jun 17 & ESPaDOnS & CFHT & 80,000 &   \\
& 2006 Jun 18 & ESPaDOnS & CFHT & 80,000 &   \\
$\upsilon$ And & 2002 Jul 27 & Gecko & CFHT & 110,000 & 13 \\
& 2002 Jul 28 & Gecko & CFHT & 110,000 &   \\
& 2002 Jul 29 & Gecko & CFHT & 110,000 &   \\
& 2002 Jul 30 & Gecko & CFHT & 110,000 &   \\
 & 2002 Aug 21 & Gecko & CFHT & 110,000 &   \\
 & 2002 Aug 23 & Gecko & CFHT & 110,000 &   \\
 & 2003 Sep 9 & Gecko & CFHT & 110,000 &   \\
 & 2003 Sep 10 & Gecko & CFHT & 110,000 &   \\
 & 2003 Sep 11 & Gecko & CFHT & 110,000 &   \\
 & 2003 Sep 12 & Gecko & CFHT & 110,000 &   \\
 & 2003 Sep 13 & Gecko & CFHT & 110,000 &   \\

\end{tabular}}

\caption{\fontfamily{phv}\selectfont{Log of observations. Note that all raw and calibrated data
are available from the CFHT (https://www.cadc-ccda.hia-iha.nrc-cnrc.gc.ca/en/cfht/)
and PolarBase (http://polarbase.irap.omp.eu/) archives.}}

\end{table}

\clearpage

\begin{figure}[htbp]
   \centering
   \includegraphics[scale=.7,clip,trim=35mm 25mm 10mm 20mm,angle=0]{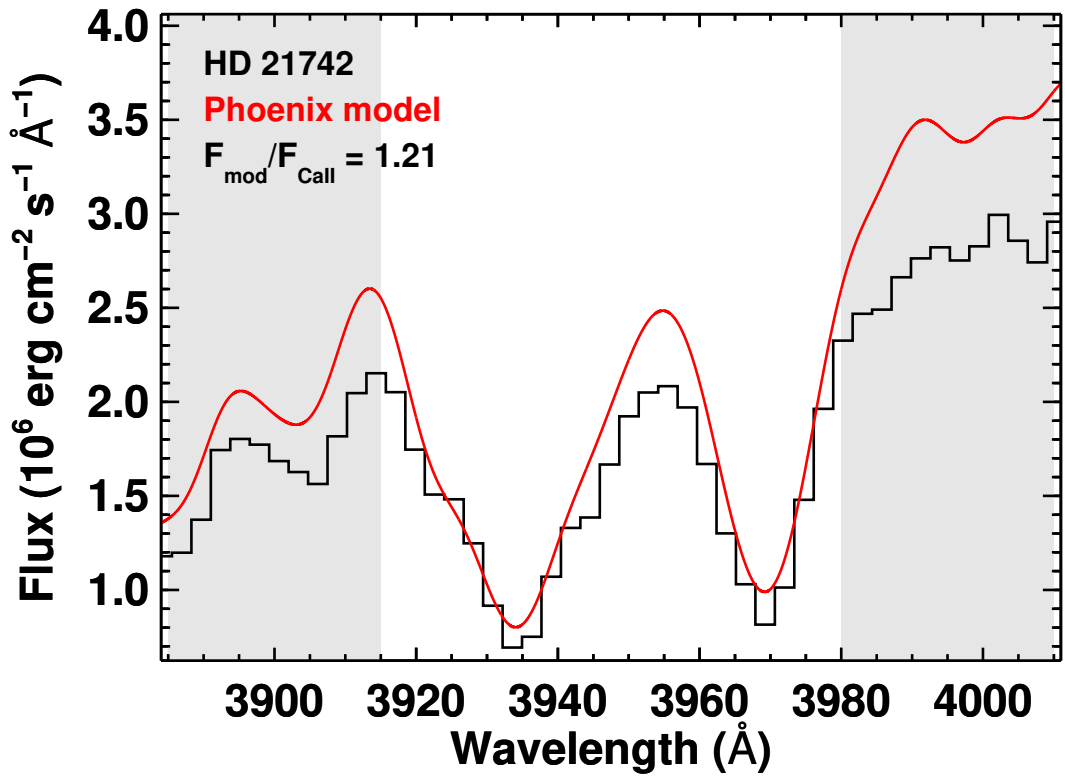}

   \caption{\fontfamily{phv}\selectfont{Example of the NGSL star HD 21742 and the PHOENIX model spectrum
   of the same $T_\text{eff}$. The value of the ratio plotted in the bottom panel of
   Supplementary Figure 2 is given in the upper-left. The continuum bands used for the mean
   flux comparison are shown with the shaded gray areas.}}

\end{figure}

\clearpage

\begin{figure}[htbp]
   \centering
   \includegraphics[scale=.75,clip,trim=10mm 60mm 10mm 35mm,angle=0]{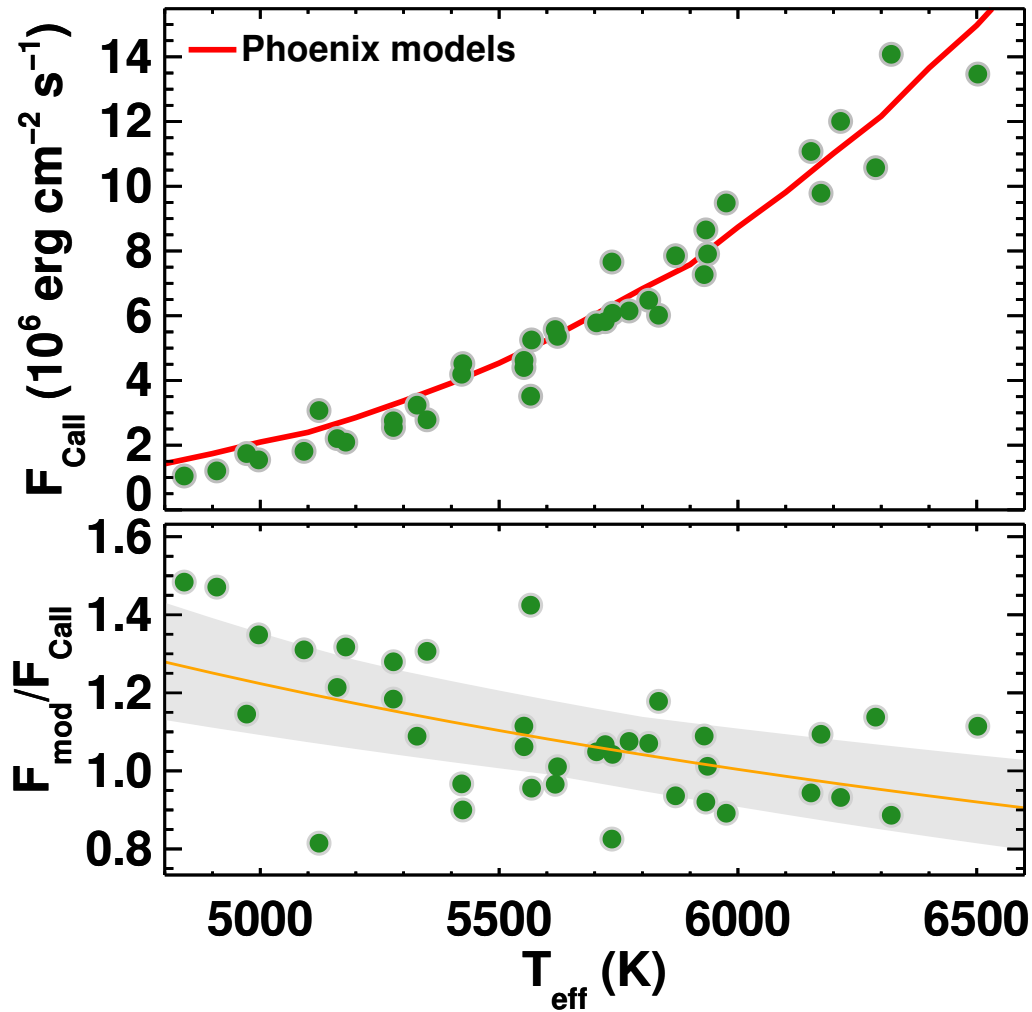}

   \caption{\fontfamily{phv}\selectfont{Top panel: surface fluxes for the selected NGSL sample (green
circles) and the flux values from the PHOENIX model spectra (solid red line).
Bottom panel: the ratio of the model fluxes to the measured fluxes. The PHOENIX
models tend to over-predict the surface fluxes for $T_\text{eff} \lesssim 5500$
K. A power-law fit to this trend is shown with the orange line, as well as the
68\% confidence intervals in the banded gray region. The power law fit is used
as a corrective factor when applying the PHOENIX models to our SPI targets.}}

\end{figure}

\clearpage

\begin{figure*}[htbp]
   \centering
   \includegraphics[scale=.7,clip,trim=5mm 5mm 15mm 5mm,angle=0]{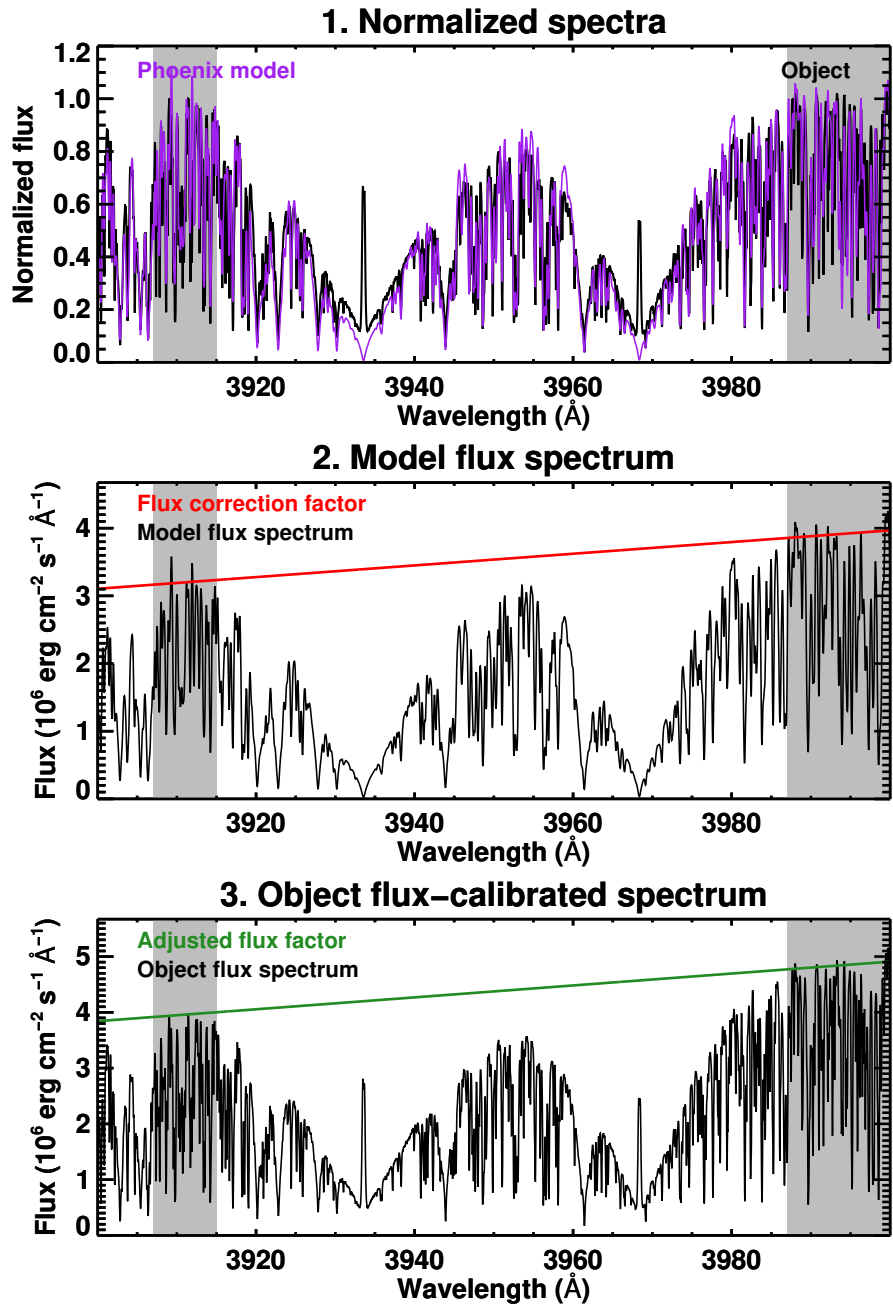}

   \caption{\fontfamily{phv}\selectfont{Step-by-step visualization of the flux calibration process. The
gray bands in all panels show the range of wavelengths used to normalize the
spectra or fit the continuum flux. The left panel shows the normalized spectrum
comparison between the object and the rotationally broadened PHOENIX model of
the same $T_\text{eff}$. The middle panel shows the linear flux continuum fit
(red line) to the PHOENIX model flux spectrum. The third panel shows the flux
continuum fit (green line) to the model spectrum, adjusted for the power law
correction shown in Supplementary Figure 2, and the flux calibrated object
spectrum.}}

\end{figure*}

} 

\end{document}